\documentclass[10pt]{article}

\begin{document}

\begin{center}
{\huge Derivation of Source-Free Maxwell and Gravitational Radiation Equations by Group Theoretical Methods}
\end{center}

\begin{center}
Jairzinho Ramos and Robert Gilmore

{\it  Physics Department, Drexel University, Philadelphia, PA 19104, USA}

email: ramos@newton.physics.drexel.edu; robert.gilmore@drexel.edu
\end{center}

\begin{abstract}
We derive source-free Maxwell-like equations in flat spacetime
for any helicity $j$ by comparing the transformation properties
of the $2(2j+1)$ states that carry the manifestly covariant
representations of the inhomogeneous Lorentz group with the
transformation properties of the two helicity $j$ states that carry
the irreducible representations of this group. The set of constraints 
so derived involves a pair of curl equations and a pair of divergence
equations. These reduce to the free-field Maxwell equations for $j=1$
and the analogous equations coupling the gravito-electric and the
gravito-magnetic fields for $j=2$.
\end{abstract}

\section{Introduction}

The linearized field equations for the Weyl curvature tensor on flat spacetime are well known to be very similar in form to Maxwell's equations, evoking an analogy referred to as gravitoelectromagnetism \cite{Maar}. Both sets of equations can be derived with the same approach based on eliminating tha gauge modes of spin 1 and spin 2 fields using Fourier analysis and group representation theory. In this way one sees the common origin of the curl and divergence operators for vector fields and 
symmetric tracefree second rank tensor fields and their intertwining to form the Maxwell combinations. Here we examine the sourcefree field equations to focus on those differential operators for arbitrary spin.

The electromagnetic field is described in two different ways:

{\it (i) Classical Formulation}.
A field is introduced having appropriate transformation properties. Not every field represents a
physically allowed state. The non-physical fields must be annihilated by appropiate equations (constraints).

{\it (ii) Hilbert space (Quantum) Formulation}.
An arbitrary superposition of states in this space represents a physically allowed state. But that
field does not have obvious transformation properties.

For the first formulation the field is required to be ``manifestly covariant.'' This requires there to
be a certain number of field components at every space-time point, or more conveniently, for every allowed
momentum vector. In the Hilbert space formulation the number of independent components is just the
allowed number of spin or helicity states.

When the number of independent field components is less than the number of components required to define the ``manifestly
covariant'' field, there are linear combinations of these components that cannot represent physically
allowed states. The function of the field equations (constraints) is to suppress these linear combinations
of components that do not correspond to physical states. Maxwell equations and the gravitational radiation equations       
fulfill this function.

Classically, the electromagnetic field is described by six field components: 
$\vec{E}(\vec{x},t)$ and $\vec{B}(\vec{x},t)$, or their components after Fourier
transformation, $\vec{E}(k)$ and $ \vec{B}(k)$, where $k$ is a 4-vector such that 
$k\cdot k=\vec{k}\cdot \vec{k}-k_{4}k_{4}=0$, where $\vec{k}$ is a 3-momentum vector and $k_{4}$      
is an energy. The quantum description involves arbitrary superposition of two helicity components for each
momentum vector. Then we have four linear combinations          
of classical field components that must be suppressed for each $k$-vector and that are annihilated by
Maxwell's equations. By comparing the transformation properties
of the basis vectors for the manifestly covariant but non-unitary representations
of the inhomogeneous Lorentz group, with the basis vectors for its
unitary irreducible but not manifestly covariant
representations, we obtain a set of constraint equations. These reduce, for $j=1$, to Maxwell's equations and,
for $j=2$, to gravitational radiation equations, both in free space.
These equations for the gravito-electric field and the gravito-magnetic field have a structure identical to the structure of Maxwell's equations.
    
The constraint equations play an important role in this derivation
since they preserve the physical states in the theory but annihilate
the non-physical ones so that the physics of massless particles can be described properly.

\section{Inhomogeneous Lorentz Group (ILG)}

The group of inhomogeneous Lorentz transformations $\{\Lambda,a\}$ has two
important subgroups.
These are the subgroup of homogeneous Lorentz transformations $\{\Lambda,0\}$  and the invariant subgroup of
translations $\{I,a\}$. Both their representations
play a role in the derivation of relativistically covariant field
equations.

\subsection{Translations $\left\{ I,a \right\}$}

All of its unitary irreducible representations
are one-dimensional, and in fact
\begin{equation}
\Gamma^k(\left\{ I, a\right\} ) = e^{ik \cdot a},
\label{a1}
\end{equation}
where $k$ and $a$ are 4-vectors and $k$ parameterizes the one-dimensional
representations.  We may define a basis state for the one-dimensional
representation $\Gamma^k$ of $\left\{ I, a \right\}$ as $| k \rangle$:
\begin{equation}
\left\{ I, a\right\} | k \rangle  = e^{ik \cdot a} | k \rangle.
\label{a2}
\end{equation}
Physically, $k$ (or $\hbar k$) has a natural interpretation as the 4-momentum of the
photon.

\subsection{Homogeneous Lorentz Transformation $\left\{ \Lambda,0 \right\}$}

We use the Minkowski metric tensor $\eta_{\mu\nu}={\rm diag}(+1,+1,+1,-1)$ for this transformation.
Any element in $SO(3,1)$ can be expressed in a $(2j+1)(2j'+1)$-dimensional 
representation $D^{jj'}$ as follows
\begin{eqnarray}
D^{jj'}(\Lambda)
&=&\exp(\vec{\theta}\cdot\vec{L}+\vec{b}\cdot\vec{B})
=\exp[(i\vec{\theta}-\vec{b})\cdot\vec{J}^{(1)}+(i\vec{\theta}+\vec{b})\cdot\vec{J}^{(2)}]
\nonumber \\
&=&D^{j}[(i\vec{\theta}-\vec{b})\cdot\vec{J}^{(1)}]D^{j'}[(\vec{i\theta}+\vec{b})\cdot\vec{J}^{(2)}] ,
\label{a3}
\end{eqnarray}
where $\Lambda$ is the 4 $\times$ 4 Lorentz transformation matrix and the operators
$\vec{L}$ generate rotations and $\vec{B}$ generate boosts. The operator
$\vec{J}^{(1)}$ has $2j+1$ dimensional representations $D^{j}$ while
$\vec{J}^{(2)}$ has $2j'+1$ dimensional representations $D^{j'}$ and
they are given by \cite{Weinberg1,Weinberg2}
\begin{displaymath}
\vec{J}^{(1)}=\frac{1}{2}(-i\vec{L}-\vec{B}) , \qquad 
\vec{J}^{(2)}=\frac{1}{2}(-i\vec{L}+\vec{B}) .
\end{displaymath}
These operators satisfy angular momentum commutation relations.

The Lie algebra $so(3,1)$ is isomorphic to the Lie algebra
for the group of $2 \times 2$ matrices $SL(2;C)$. There are two
isomorphisms which give rise to the two matrix representations: $D^{j0}:\vec{B}=i\vec{L} \rightarrow \vec{J}^{(j)}$ and
 $D^{0j'}:\vec{B}=-i\vec{L} \rightarrow \vec{J}^{(j')}$ for any $j$.
Here $\vec{J}^{(j)}$ are the $(2j+1)\times (2j+1)$ angular momentum
matrices and $\vec{L}=i\vec{J}^{(j)}$ which will be useful later.

The representations of the connected component of the homogeneous Lorentz group $SO(3,1)$ are
described by two angular momentum indices $j$ and $j'$ which describe
each of the two representations $\vec{J}^{(1)}$ and
$\vec{J}^{(2)}$ respectively. The most general representation of the group $SO(3,1)$ is 
\begin{equation}
D^{jj'}(\vec{\theta}\cdot\vec{L} + \vec{b}\cdot\vec{B}) 
=
D^{j0}[\vec{\theta}\cdot\vec{L} + \vec{b}\cdot\vec{B}]
D^{0j'}[\vec{\theta}\cdot\vec{L} + \vec{b}\cdot\vec{B}] .
\label{a4}
\end{equation}
The first representation $D^{j0}$ is obtained using the first
isomorphism in (\ref{a3}). In a similar way we obtain the second
representation $D^{0j'}$. Thus, (\ref{a4}) becomes
\begin{equation}
D^{jj'}(\Lambda)
=D^{jj'}(\vec{\theta}\cdot\vec{L} + \vec{b}\cdot\vec{B})
=
\exp[(i\vec{\theta}-\vec{b})\cdot J^{(j)}]
\exp[(i\vec{\theta}+\vec{b})\cdot J^{(j')}] . 
\label{a5}
\end{equation}
The action of $\Lambda$ on the basis $\tiny{\mid \begin{array}{cc}
j & j' \\
m & m'
\end{array}
\rangle}$ of the total space $\xi(j,j')$ through the representation
$D^{jj'}(\Lambda)$ is expressed by
\begin{equation}
\Lambda \mid
\begin{array}{cc}
j & j' \\
m & m'
\end{array}
\rangle=\mid
\begin{array}{cc}
j & j' \\
l & l'
\end{array}
\rangle D^{jj'}_{ll',mm'}(\Lambda) .
\label{a6}
\end{equation}

\subsection{Representations of the Group $O(3,1)$}

The full homogeneous Lorentz group $O(3,1)$ \cite{Heine1,Heine2} consists of those operations that
are connected to the identity, that are in the subgroup $SO(3,1)$,
together with the parity and time reversal operations $P$ and $T$ and their product $PT=TP$.
The action of these operations on the coordinates are:
\begin{equation}
P(x,y,z,ct) = (-x,-y,-z,+ct) , \qquad
T(x,y,z,ct) = (+x,+y,+z,-ct) .
\label{a61}
\end{equation}
The action of these discrete operators \cite{Weinberg3} on the generators of
infinitesimal transformations 
$\vec{L}=i\vec{J}$ $(L_i = \epsilon_{ijk}x^j\partial_k)$ and 
$\vec{B}=i\vec{K}$ $(B_i=-x^i\partial_4 - x^4 \partial_i)$ are
\begin{eqnarray}
P  \vec{L} P^{-1} = \vec{L} ,
\,\,\,\,\,\,\,\,\,\,\,\  
P \vec{B} P^{-1} = -\vec{B} ,
\nonumber \\
T \vec{L} T^{-1} = \vec{L} ,
\,\,\,\,\,\,\,\,\,\,\,\ 
T \vec{B} T^{-1} = -\vec{B} . 
\label{a62}
\end{eqnarray}
The full homogeneous Lorentz group consists of four
components, each similar to $SO(3,1)$.  The full group can be written as
follows:
\begin{equation}
O(3,1) = \left\{ I, P, T, PT \right\} \otimes SO(3,1) .
\label{a63}
\end{equation}
The four operators $\left\{ I, P, T, PT \right\}$ form a discrete
invariant subgroup of $O(3,1)$ and act as coset representatives for
the quotient group $O(3,1)/SO(3,1)$.  The four components of the group
$O(3,1)$ are connected to the group operations: $I, P, T, PT$.

The discrete operations $P, T$ interchange the operators
$\vec{J}^{(1)}$ and $\vec{J}^{(2)}$.  As a result, the matrices
$D^{jj'}=D^{j0}\otimes D^{0j'}$ can represent the connected component of the homogeneous
Lorentz group $SO(3,1)$, but cannot represent the full homogeneous
Lorentz group $O(3,1)$ unless $j=j'$.  If $j \ne j'$ the matrix
representation of the full group consists of the direct sum of the two
matrix representations $D^{jj'} \oplus D^{j'j}$ where $D^{j'j}=D^{j'0}\otimes D^{0j}$.

In the following sections we will be particularly interested in
describing massless particles with helicity $j$ (integer)
in terms of the representations $D^{j0} \oplus D^{0j}$
or $D^{\frac{j}{2}\frac{j}{2}} = D^{\frac{j}{2}0} \otimes D^{0\frac{j}{2}}$.  
The former class of representations act on states (fields)
with $2(2j+1)$ components while the latter act on states
(potentials) with
$(j+1)^2$ components.  For the photon with helicity $j=1$ there
are six components (the three components of the electric field
and the magnetic field, $\vec{E}$ and $\vec{B}$), and four
components (of the vector potential $A_{\mu}$), respectively.
These observations are usefully summarized:
\begin{displaymath}  
\begin{array}{ccccc}
{\rm Particle} & &{\rm Photon} & & {\rm Graviton} \\
  j &\hspace{0.5cm} & 1 &\hspace{1.0cm} & 2 \\
D^{j0} \oplus D^{0j} & & \vec{E}, ~\vec{B} & &  Q_{E}, ~Q_{B} \\
D^{\frac{j}{2} 0} \otimes D^{0 \frac{j}{2}} & & A_{\mu} & & h_{\mu\nu}
\end{array}
\end{displaymath}
For the photon, description of the electromagnetic field is in terms
of the fields $\vec{E}, ~\vec{B}$ which carry the representation
$D^{j0} \oplus D^{0j}$ with $j=1$, or in terms of the vector
potential $A_{\mu}$, which carries the representation
$D^{\frac{j}{2}\frac{j}{2}} = D^{\frac{j}{2} 0} \otimes D^{0 \frac{j}{2}}$, also with $j=1$.  
The fields $\vec{E}, ~\vec{B}$ are tracefree dipole fields.

For the graviton, description of the gravitational field is in terms
of the fields $Q_{E}, ~Q_{B}$ which carry the representation
$D^{j0} \oplus D^{0j}$ with $j = 2$, or in terms of the linear metric perturbation
$h_{\mu\nu}$, which is a symmetric tracefree tensor representing the gravitational potential and carries the representation 
$D^{\frac{j}{2}\frac{j}{2}}$, also with $j = 2$.  The fields $Q_{E}, ~Q_{B}$
are tracefree quadrupole fields, each with five components.

\section{Representations of the ILG}

We construct two kinds of representations for the ILG. These are the
manifestly covariant representations and the unitary irreducible
representations \cite{Wigner}.

\subsection{Manifestly Covariant Representations}

We construct manifestly covariant representation of the ILG by
constructing direct products of the basis vectors 
\begin{equation}
|k \rangle \otimes \mid
\begin{array}{cc}
j & j' \\
m & m'
\end{array}
\rangle=|k \rangle \mid
\begin{array}{cc}
j & j' \\
m & m'
\end{array}\rangle 
\label{a7}
\end{equation}
for the subgroups of the ILG.

The action of the ILG on these direct product bases (\ref{a7}) is defined by
the action of the two subgroups, homogeneous Lorentz transformation and translations, on the momentum basis states $|k \rangle$
and field component basis states $ \tiny{\mid
\begin{array}{cc}
j & j' \\
m & m'
\end{array}
\rangle}$ separately:

\begin{itemize}
\item The action of the $\{I,a\}$ on these basis states is defined by
\begin{equation}
\{I,a\}|k \rangle=|k \rangle e^{ik\cdot a} ,
\label{a8}
\end{equation}
\begin{equation}
\{I,a\}\mid
\begin{array}{cc}
j & j' \\
m & m'
\end{array}\rangle
=
\mid
\begin{array}{cc}
j & j' \\
l & l'
\end{array}\rangle \delta_{ll',mm'} .
\label{a9}
\end{equation}
\item The action of $\{\Lambda,0\}$ on these basis states is defined, from (\ref{a8}) and (\ref{a6}) respectively, by
\begin{equation}
\{\Lambda,0\}|k \rangle=|\Lambda k \rangle ,
\label{a10}
\end{equation}
\begin{equation}
\{\Lambda,0\}\mid
\begin{array}{cc}
j & j' \\
m & m'
\end{array} \rangle=\mid
\begin{array}{cc}
j & j' \\
l & l'
\end{array} \rangle D^{jj'}_{ll',mm'}(\Lambda) .
\label{a11}
\end{equation}
\end{itemize}

\subsection{Unitary Irreducible Representations}

We have a representation of $\{\Lambda,a\}$ that is unitary and
irreducible. For the subgroup $\{I,a\}$ this reduces to a direct sum
of irreducibles $\Gamma^{k}(\{I,a\})$. The basis states are
$|k;\xi\rangle$, so that
\begin{equation}
\{I,a\}|k;\xi\rangle = |k;\xi\rangle e^{ik\cdot a},
\label{a12}
\end{equation}
where $\xi$ is a helicity for different states with the same
4-momentum.

The action of the subgroup $\{\Lambda,0\}$, by using (\ref{a12}) and
the invariance of the inner product, on the basis states $|k;\xi\rangle$
is defined by
\begin{equation}
\{\Lambda,0\}|k;\xi \rangle=|\Lambda k;\xi'\rangle M_{\xi' \xi}(\Lambda) ,
\label{a13}
\end{equation}
where the matrix $M_{\xi' \xi}(\Lambda)$ remains to be determined.

We consider only the case of zero mass particles in the present
discussion. To construct the matrix $M(\Lambda)$, we
choose one particular 4-vector $k^{0}$ for this case
\begin{displaymath}
\,\,\,\,\,\,\,\,\,\,\,\,\,\,\,\,\
\begin{array}{ccccc} 
 k\cdot k=0 , & k \ne 0 ,& k^0=(0,0,1,1)   
\end{array}
\end{displaymath}
\typeout{!!! there are two assignments to $k^0$ but you talk about one particular 4-vector !!!}%
where $T$ is the time reversal operator and the vector $k^{0}$ is
called the {\bf little vector} \cite{Wigner}.

The effect of a homogeneous Lorentz transformation on the state
$|k^0; \xi \rangle$ is determined by writing  $\Lambda$ as a
product of two group operations
\begin{equation}
\Lambda = C_{k}H_{k^0} ,
\label{a14}
\end{equation}
where $H_{k^0}$ is the stability subgroup (little group [7-9]) of the little vector
$k^0$
\begin{equation}
H_{k^0} k^0=k^0 .
\label{a15}
\end{equation}
For our case the little group is $ISO(2)$, and $C_{k}$ is a coset representative 
\begin{equation}
C_{k} k^0=k=\Lambda k^{0} .
\label{a16}
\end{equation}
The generators of the little group $H_{k^0}$ are defined by
\begin{equation}
H_{k^{0}}=I_{4}+\alpha G,
\label{a17}
\end{equation}
and the generator $G$ is given by
\begin{eqnarray}
\tiny{
G = \left(
\begin{array}{cccc}
0 & \theta_{3} & -\theta_{2} & -b_{1} \\
-\theta_{3} & 0 & \theta_{1} & -b_{2} \\
\theta_{2} & -\theta_{1} & 0 & -b_{3} \\
-b_{1} & -b_{2} & -b_{3} & 0
\end{array}
\right)}=\vec{\theta}\cdot\vec{L}+\vec{b}\cdot\vec{B} .
\label{a18}
\end{eqnarray}
From (\ref{a15}), an arbitrary element in this Lie subgroup
$H_{k^0}$ acting on $k^{0}$ must leave $k^{0}$ invariant, therefore
\begin{equation}
Gk^{0}=0 .
\label{a19}
\end{equation}
Consequently, the stability subalgebra is defined by: 
$b_{3}=0$, $b_{2}=\theta_{1}$, $b_{1}=-\theta_{2}$.
The generator $G$ in this subalgebra becomes
\begin{equation}
G=G_{H}=\theta_{1}Y_{1}+\theta_{2}Y_{2}+\theta_{3}Y_{3} ,
\label{a20}
\end{equation}
where $Y_{1}=L_{1}+B_{2}$, $Y_{2}=L_{2}-B_{1}$, $Y_{3}=L_{3}$. These
operators satisfy the commutation relations for the group $ISO(2)$
(inhomogeneous motions of the Euclidean plane $R^{2}$).

The action of little group on the subspace of states
$|k^{0};\xi\rangle$ is
\begin{equation}
H_{k^{0}}|k^{0};\xi \rangle=|H_{k^{0}}k^{0};\xi' \rangle D_{\xi' \xi}(H_{k^{0}})=|k^{0};\xi' \rangle D_{\xi' \xi}(H_{k^{0}}) .
\label{a21}
\end{equation}
The original representation of the ILG is unitary and irreducible if
and only if the representation $D_{\xi'\xi}(H_{k^{0}})$ of the little
group is unitary and irreducible.

We construct the unitary and irreducible representation of $ISO(2)$
following the method of little groups. Since $ISO(2)$ has a
two-dimensional translation invariant group, basis states in a unitary
irreducible representation can be labeled by a vector
$\kappa=(\kappa_{1},\kappa_{2})$ in a 2-dimensional Euclidean space,
$\kappa \in R^{2}$, $\kappa\cdot\kappa \geq 0$. The invariant length of
$\kappa$ parameterizes the representation.

Physically, we require $\kappa=0$ [7-9]. With
this, the physically allowable representation of the little group
$(\theta_{1}\rightarrow 0, \theta_{2}\rightarrow 0)$ \cite{Weinberg4} is
\begin{equation}
\,\,\,\,\,\,\,\,\,\,\,\,\,\,\,\
D_{\xi'\xi}(H_{k^{0}})
=\exp (\theta_{1}Y_{1}+\theta_{2}Y_{2}+\theta_{3}Y_{3})
=\exp(i\theta_{3}J_{3})
=\exp (i\xi\theta_{3})\delta_{\xi'\xi} ,
\label{a22}
\end{equation}
where $J_{3}$ is the $(2j+1)$-dimensional angular momentum diagonal
matrix and $\xi$ is integer or half-integer($-j \leq \xi \leq j$).

The action of an arbitrary element of the ILG on any state in this
Hilbert space is
\begin{displaymath}
\{\Lambda,a\}|k;\xi\rangle =\{\Lambda,0\}\{I,\Lambda^{-1}a\}|k;\xi\rangle.
\end{displaymath}
By using the coset representative $C_{k}$ that permutes the 4-vector
subspaces: $C_{k}|k^{0};\xi\rangle=|k;\xi\rangle$, and (\ref{a22}) the last equation
becomes
\begin{equation}
\{\Lambda,a\}|k;\xi\rangle=|k';\xi\rangle e^{i\xi\Theta}e^{i\Lambda k\cdot a},
\label{a23}
\end{equation}
where
\begin{displaymath}
H_{k^{0}}=C^{-1}_{k'}\Lambda C_{k} \,\ \longrightarrow \,\ e^{i\xi\Theta} .
\end{displaymath}
Thus, we have obtained the action of the homogeneous and the ILG on
a state $|k,\xi \rangle$ in the Hilbert space in terms of their
unitary irreducible representation by using the little group $H_{k^{0}}$. It will be
useful to compare this quantum description with the classical one.

\section{Transformation Properties}

To compare these two kinds of representations for massless particles
we compare transformation properties of their states in Table \ref{table1}, where $\theta_{\pm}=\theta_{1}\pm i\theta_{2}$ and $J_{\pm}=J_{1}\pm iJ_{2}$.
\begin{table}
\caption{Manifestly covariant versus unitary irreducible representations.}
\begin{displaymath}
\begin{array}{c|c}
 Classical & Quantum \\
 & \\
{\bf Manifest.~Covariant.~Rep.} &  {\bf Unit.~Irreducible.~Rep.} \\
 {\rm basis~state:\ } |k\rangle \tiny{\mid \begin{array}{cc} j & j' \\ m & m' \end{array}\rangle}, k=\Lambda k^{0} & 
{\rm basis~state:\ } |k;\xi\rangle, k=\Lambda k^{0} \\ \hline
 &  \\
\{H_{k^{0}},0\}|k^{0}\rangle \tiny{\mid \begin{array}{cc} j & j' \\ m & m' \end{array}\rangle}=
|k^{0}\rangle \tiny{\mid \begin{array}{cc} j & j' \\ l & l' \end{array}\rangle}D^{jj'}_{ll',mm'}(H_{k^{0}}) &  \{H_{k^{0}},0\}|k^{0};\xi\rangle=
|k^{0};\xi\rangle e^{i\xi\Theta}\,\,\,\ {\bf (c)} \\
 & \\
{\rm for}\,\ j'=0 \,\ \vec{B}=i\vec{L} & \\
D^{j0}(H_{k^{0}})=\exp [i\theta_{3}J^{(j)}_{3}+i\theta_{-}J^{(j)}_{+}] & \\
 & \\
{\rm if}\,\ j=m=\xi>0 & \\
\{H_{k^{0}},0\}|k^{0}\rangle \tiny{\mid \begin{array}{cc} j & 0 \\ j & 0 \end{array}\rangle} \rightarrow 
|k^{0}\rangle \tiny{\mid \begin{array}{cc} j & 0 \\ j & 0 \end{array}\rangle}e^{i\xi\theta_{3}}\,\,\,\ {\bf (a)} & \\
 & e^{i\xi\Theta}=\exp (\Theta Y_{3}+\theta_{1}Y_{1}+\theta_{2}Y_{2})\\
{\rm for}\,\ j=0 \,\ \vec{B}=-i\vec{L} & \\
D^{0j'}(H_{k^{0}})=\exp [i\theta_{3}J^{(j')}_{3}+i\theta_{+}J^{(j')}_{-}] & \\
 & \\
{\rm if}\,\ j'=-m'=-\xi, \xi<0 & \\
\{H_{k^{0}},0\}|k^{0}\rangle \tiny{\mid \begin{array}{cc} 0 & j' \\ 0 & -j' \end{array}\rangle} \rightarrow 
|k^{0}\rangle \tiny{\mid \begin{array}{cc} 0 & j' \\ 0 & -j' \end{array}\rangle}e^{i\xi\theta_{3}}\,\,\,\ {\bf (b)} &  
\end{array}
\end{displaymath}
\label{table1}
\end{table}
By comparing {\bf (a)} and  {\bf (b)} with  {\bf (c)} we conclude that
the state $|k^{0}\rangle \tiny{\mid \begin{array}{cc} j & 0 \\ j & 0
  \end{array}\rangle}$ transforms identically to $|k^{0};\xi\rangle$ if $j=\xi$ and $\xi>0$. 
The state $|k^{0}\rangle \tiny{\mid \begin{array}{cc} 0 & j' \\ 0 & -j'
  \end{array}\rangle}$ transforms identically to $|k^{0};\xi\rangle$ if $j'=-\xi$ and $\xi<0$.

Therefore, when $j=\xi>0$, the state $|k^{0}\rangle \tiny{\mid \begin{array}{cc} j & 0 \\ j & 0 \end{array}\rangle}$
is the unique physical state in the manifestly covariant
representation. The remaining states are superfluous (non-physical
states). Thus, the amplitudes $\langle k^{0}; \tiny{\begin{array}{cc} j  & 0 \\ m & 0 \end{array}}|\psi \rangle$ of the states
$|k^{0} \rangle \tiny{\mid \begin{array}{cc} j  & 0 \\ m & 0 \end{array} \rangle} $, with $m \neq j$, must all vanish because they are
all non-physical states which are required in the manifestly covariant
representation but are not present in the Hilbert space that carries the unitary irreducible representation.

A simple linear way to enforce this condition on the non-physical
amplitudes is to require
\begin{equation}
k^{0}_{3}\{J^{(j)}_{3}-jI_{2j+1}\} \langle k^{0}; \begin{array}{cc} j & 0 \\ m & 0 \end{array}|\psi \rangle=0 .
\label{a24}
\end{equation}
The matrix within the bracket $\{\}$ is diagonal, with the elements $(j-j)k^{0}_{3}=0$
multiplying the physically allowed amplitude $\langle k^{0};\tiny{
\begin{array}{cc} j & 0 \\ j & 0 \end{array}}|\psi \rangle$. Therefore
this amplitude is arbitrary. The non-zero elements $(m-j)k^{0}_{3}$ multiplying the non-physical amplitudes
$\langle k^{0};\tiny{\begin{array}{cc} j & 0 \\ m & 0
\end{array}}|\psi \rangle$, $m \neq j$, absent in the description of a
physical state, require that these amplitudes be null. For $\xi<0$, by the same argument, we obtain an equation similar to (\ref{a24}).

By using the coset operator $C_{k}$: $C_{k}|k^{0};\xi \rangle=|k;\xi
\rangle$ and $C_{k}|k^{0} \rangle \mid \tiny{\begin{array}{cc} j & j' \\ m & m' \end{array}}\rangle=
|k \rangle \tiny{\mid \begin{array}{cc} j & j' \\ l & l' \end{array}\rangle}
D^{jj'}_{ll',mm'}(C_{k})$ we can express the constraint equation
(\ref{a24}) for any 
$k=(k_{1},k_{2},k_{3},k_{4})$, $k^{2}_{1}+k^{2}_{2}+k^{2}_{3}=k^{2}_{4}$.
The condition on the amplitudes in the subspace $|k\rangle$ is related
to condition in the subspace $|k^{0}\rangle$ by a similarity
transformation. Considering $C_{k}$ as the product of a boost
$B_{z}$ in the $z$ direction followed by a rotation $R(\vec{k})$, the
constraint equation for any $k$ and $\xi>0$ becomes
\begin{equation}
(\vec{J}^{(j)}\cdot\vec{k}-jk_{4}I_{2j+1})\langle k; \begin{array}{cc} j &  0 \\ m & 0 \end{array}|\psi \rangle=0 ,
\label{a25}
\end{equation}
where $\vec{J}^{(j)}$ are the three $(2j+1)\times (2j+1)$ angular
momentum matrices. For any $\xi<0$, we obtain a similar constraint
equation.

\section{The Constraint Equation}
 
The constraint equation is conveniently expressed in the coordinate
rather than the momentum representation by inverting the original
Fourier transform that brought us from the coordinate to the momentum representation.
Since $e^{ik\cdot x} = e^{ik_{\mu} x^{\mu}} = $ $e^{i(\vec{k}\cdot\vec{x} - k_4ct)}$,
we can replace $\vec{k} \rightarrow \frac{1}{i}\nabla$
and $k_4 \rightarrow -\frac{1}{i}\frac{1}{c}\frac{\partial}{\partial t}$. The Fourier inversion is explicitly 
\begin{equation}
\langle k | x \rangle
\left\{\vec{J}^{(j)}.\frac{1}{i} \nabla + j
\frac{1}{i} \frac{\partial }{\partial (ct)} I_{2j+1} \right\}
\langle x | k \rangle \langle k;
\begin{array}{cc} j & 0 \\ m & 0 \end{array} | \psi \rangle =0 .
\label{a26}
\end{equation}
We define complex spherical tensor fields
\begin{equation}
\psi_{jm}=\langle x | k \rangle \langle k;
\begin{array}{cc} j & 0 \\ m & 0 \end{array} | \psi \rangle=T_{E}^{(j)}(x)+iT_{B}^{(j)}(x) ,
\label{a27}
\end{equation}
where $-j \leq m \leq j$; $x=(\vec{x},t)$ and $T_{E}^{(j)}$ and
$T_{B}^{(j)}$ are a spherical tensor and pseudo-tensor respectively, of rank $j$ with $2j+1$ components each. Thus, in
coordinate space, the constraint equation becomes
\begin{equation}
\left\{ \frac{1}{i}\frac{\vec{J}^{(j)}.\nabla}{j} +
\frac{1}{i}\frac{\partial }{\partial (ct)} I_{2j+1} \right\}
\left(T_{E}^{(j)} + iT_{B}^{(j)} \right)=0 .
\label{a28}
\end{equation}
We define the curl operator in a new way \cite{Gilmore}: 
$curl =\frac{1}{i}\frac{\vec{J}^{(j)}.\nabla}{j}$, which is equivalent to the standard definition of the $curl$ operator in three dimensions. 
With this definition we obtain
\begin{equation}
\left\{curl -
\frac{i}{c}\frac{\partial }{\partial t} I_{2j+1} \right\}
\left(T_{E}^{(j)} + iT_{B}^{(j)} \right)=0.
\label{a29}
\end{equation}
Finally, we take the real and imaginary parts of this equation and
obtain:
\begin{eqnarray}
curl T_{E}^{(j)}+\frac{1}{c}\frac{\partial}{\partial t}T_{B}^{(j)}=0 \nonumber \\
curl T_{B}^{(j)}-\frac{1}{c}\frac{\partial}{\partial t}T_{E}^{(j)}=0.
\label{a30}
\end{eqnarray}
Thus we conclude that the zero mass particles with helicity $j$ can
be described, at least in free space, by a pair of real coupled,
interacting, oscillating spherical tensor fields with $(2j+1)$
components each.

\subsection{Symmetries of the Spherical Tensor Fields}

For integer $j$, the fields
$T_{E}^{(j)}$ and $T_{B}^{(j)}$ are tracefree rank-$j$
tensor and pseudo-tensor fields, respectively.  They transform under
$D^{(j)}(SO(3))$ of the proper rotation group $SO(3)$,
and each has $(2j+1)$ components.  Their
transformation properties under the discrete group operations $P$ and
$T$ are summarized in this table:
\begin{displaymath}  
\begin{array}{c|cc}
{\rm Discrete} & T_{E}^{(j)} & T_{B}^{(j)}  \\
{\rm Operation} & & \\   \hline
P & (-)^j & -(-)^j \\
T & -(-)^j & (-)^j 
\end{array}
\end{displaymath}
Under these transformation properties, the curl equations
are invariant under the subgroup of discrete operations:
$\{ I, P, T, PT \}$.

\section{The Source Free Equations}

Eq.~(\ref{a30}), or (\ref{a29}), describes the source-free Maxwell
equations for $j=1$ and the gravitational radiation equations for $j=2$ in the weak field
limit. We exhibit these equations below.

\subsection{Maxwell's Equations}

In Cartesian coordinates the equation (\ref{a29}) reduces, for $j=1$,
to
\begin{equation}
\left(\begin{array}{ccc}
-\frac{i}{c}\frac{\partial}{\partial t} & -\partial_{z} & \partial_{y} \\
\partial_{z} & -\frac{i}{c}\frac{\partial}{\partial t} & -\partial_{x} \\
-\partial_{y} & \partial_{x} & -\frac{i}{c}\frac{\partial}{\partial t} 
\end{array}\right)
\left(\begin{array}{c}
E_{x}+iB_{x} \\
E_{y}+iB_{y} \\
E_{z}+iB_{z}
\end{array}\right)=0 .
\label{a31}
\end{equation}
These three equations are expressed as a vector equation by
\begin{equation}
-\frac{i}{c}\frac{\partial }{\partial t}(\vec{E}+i\vec{B})
+ {\nabla \times} (\vec{E}+i\vec{B})=0 .
\label{a32}
\end{equation}
The real and imaginary parts of these equations are equal to zero,
(\ref{a30}), so we obtain
\begin{eqnarray}
\nabla \times \vec{E} + \frac{1}{c}\frac{\partial
\vec{B}}{\partial t} &=& 0 \nonumber \\
\nabla \times \vec{B} - \frac{1}{c}\frac{\partial
\vec{E}}{\partial t} &=& 0.
\label{a33}
\end{eqnarray}
These are the two Maxwell ``curl'' equations.

The two Maxwell ``div'' equations are also present.  Eq.~(\ref{a25}) that forces the non-physical components of the state
$|\tiny{\begin{array}{cc} 1 & 0 \\ m & 0 \end{array} \rangle}$ to
vanish says that, in the special frame with little vector
$k^{0}=(0,0,1,1)$, the only nonvanishing component is the component
with $m=j=1$.  The coordinates of this component are $-(x+iy)$ and the
spatial part of the $k$ vector is $(0,0,1)$. In this frame
$x$ and $y$ coordinates are arbitrary.  In general, this states that 
$\vec{k}\cdot (\vec{E} + i\vec{B})=0$.
With the substitution $\vec{k} \rightarrow \frac{1}{i} \nabla$,
the real and imaginary parts of this complex equation reduce to
\begin{eqnarray}
\nabla\cdot\vec{B}=0 ,   
\nonumber \\
\nabla\cdot\vec{E}=0 .
\label{a34}
\end{eqnarray}
Thus, we obtain the four Maxwell's equations for the electromagnetic
field (in vacuum). The four equations (\ref{a33}) and (\ref{a34}) are
necessary to obtain the source-free wave equation for these
fields: $(\nabla^{2}-\frac{1}{c^{2}}\frac{\partial^{2}}{\partial t^{2}})\vec{E}=0$ and 
$(\nabla^{2}-\frac{1}{c^{2}}\frac{\partial^{2}}{\partial t^{2}})\vec{B}=0$.

\subsection{Gravitational Radiation Equations}

For $j=2$, equations (\ref{a30}) describe two interacting tracefree
quadrupole fields $Q_{E}$ and $Q_{B}$ (tensors of rank
two). Specifically, the equations are
\begin{eqnarray}
curl Q_{E}+\frac{1}{c}\frac{\partial Q_{B}}{\partial t}=0 ,
\nonumber \\
curl Q_{B}-\frac{1}{c}\frac{\partial Q_{E}}{\partial t}=0 .
\label{a35}
\end{eqnarray}

In Cartesian coordinates these coupled interaction equations, for $Q_{E}=F$ and $Q_{B}=G$, become
\begin{eqnarray} \label{a36}
\,\,\,\,\,\,\,\,\,\,\,\,\,\,\,\,\,\,\,\,\
\tiny{\frac{1}{2}
\left(\begin{array}{ccccc}\
0 & \partial_y & -\partial_x & 2\partial_z & 0 \\
-\partial_y & 0 & \partial_z & -\partial_x & -\sqrt{3} \partial_x \\
\partial_x & -\partial_z & 0 & -\partial_y & \sqrt{3}\partial_y \\
-2\partial_z & \partial_x & \partial_y & 0 & 0 \\
0 & \sqrt{3} \partial_x & -\sqrt{3}\partial_y & 0 & 0 \end{array}
\right)
\left(  \begin{array}{c}F_1 \\ F_2 \\ F_3 \\ F_4 \\ F_5 \end{array}
\right)
+ \frac{1}{c}\frac{\partial }{\partial t}
\left(  \begin{array}{c}G_1 \\ G_2 \\ G_3 \\ G_4 \\ G_5 \end{array}
\right)=0}  ,
\nonumber \\
& & \label{eq:39}\\
\,\,\,\,\,\,\,\,\,\,\,\,\,\,\,\,\,\,\,\,\
\tiny{\frac{1}{2}
\left(  \begin{array}{ccccc}
0 & \partial_y & -\partial_x & 2\partial_z & 0 \\
-\partial_y & 0 & \partial_z & -\partial_x & -\sqrt{3} \partial_x \\
\partial_x & -\partial_z & 0 & -\partial_y & \sqrt{3}\partial_y \\
-2\partial_z & \partial_x & \partial_y & 0 & 0 \\
0 & \sqrt{3} \partial_x & -\sqrt{3}\partial_y & 0 & 0 \end{array}
\right)
\left(  \begin{array}{c}G_1 \\ G_2 \\ G_3 \\ G_4 \\ G_5 \end{array}
\right)
- \frac{1}{c}\frac{\partial }{\partial t}
\left(  \begin{array}{c}F_1 \\ F_2 \\ F_3 \\ F_4 \\ F_5 \end{array}
\right)=0}. \nonumber
\end{eqnarray}
There is another pair of equations for gravitational radiation, in
complete analogy with the ``divergence'' equations of
electromagnetism.  In the frame with $k^{0} = (0,0,1,1)$ the
state $\tiny{|\begin{array}{cc} 2 & 0 \\ m & 0 \end{array}
\rangle}$ vanishes unless $m=2$. This state has Cartesian coordinates
$(x+iy)^2 = (x^2-y^2)+2ixy$.  All other components in Cartesian
coordinates: $yz$, $zx$, and $(2z^2-x^2-y^2)$ are forced to vanish.  
In an arbitrary coordinate system, where the spatial
component of the $k$ vector is $\vec{k}$, this condition is
$k^i S_{ij} = 0$ and this becomes, after the standard Fourier
transform, $\partial_i S_{ij}=0$.  Here $S_{ij}$ is a tracefree second
order symmetric tensor on $R^3$ obeying $S_{ij}=S_{ji}$ and
$S_{ii}=0$.  The tracefree tensor $S_{ij}$ and the five-components
of $Q_E=F$ and $Q_B=G$ are related as follows:

For $F$ 
\begin{displaymath}
\,\,\,\,\,\,\
F_1=F_{xy} ,\
F_2=F_{yz} ,\ 
F_3=F_{zx} ,\ 
F_4=F_{\frac{1}{2}(x^2-y^2)} ,\ 
F_5=F_{\frac{1}{2\sqrt{3}}(2z^2-x^2-y^2)} ,  
\end{displaymath}
\begin{equation}
\tiny{S_{ij}=\left(\begin{array}{ccc}
F_{11} & F_{12} & F_{13} \\
F_{21} & F_{22} & F_{23} \\
F_{31} & F_{32} & F_{33} \end{array} \right)=
\left(\begin{array}{ccc}
F_4-\frac{1}{\sqrt{3}}F_5 & F_1 & F_3 \\
F_1 & -F_4-\frac{1}{\sqrt{3}}F_5 & F_2  \\
F_3 & F_2 & \frac{2}{\sqrt{3}}F_5 \end{array} \right)} .
\label{a37}
\end{equation}
For $G$, the relation with $S_{ij}$ is similar to
(\ref{a37}). Therefore the divergence equations for gravitational
radiation, for the fields $F$ and $G$ in the representation given
above, are 
\begin{eqnarray}
\frac{\partial}{\partial x_{i}}F_{ij}&=&0 ,
\nonumber \\
\frac{\partial}{\partial x_{i}}G_{ij}&=&0 ,
\label{a38}
\end{eqnarray}
where $i,j=1,2,3$.

The decomposition of Weyl tensor into electric and magnetic tensors  $E_{ab}$ and $H_{ab}$ respectively has a long history [11-13].
In terms of these tensors the gravitational radiation equations (no sources)  is described \cite{Maar} by
\begin{equation}
\begin{array}{c}
\displaystyle (\nabla \times E)_{ab}+\frac{1}{c}\frac{\partial H_{ab}}{\partial t}=0 , \\
 \\
\displaystyle (\nabla \times H)_{ab}-\frac{1}{c}\frac{\partial E_{ab}}{\partial t}=0 , \\
\end{array}
\label{b1}
\end{equation}
where \cite{Maar2,Maar3}
\begin{displaymath}
E_{ab}=C_{acbd}u^{c}u^{d} ,\qquad 
H_{ab}=\frac{1}{2}\eta_{acde}C^{cd}_{\,\,\,\ bf}u^{e}u^{f} ,
\end{displaymath}
with $u^{a}$ a four-velocity and $C_{acbd}$ is the Weyl Tensor
\cite{Weinberg}. In four dimensions and in terms of the Riemann Tensor
$R_{abcd}$, the Ricci tensor $R_{ac}$ and the scalar curvature $R$,
the Weyl tensor can be written as follows
\begin{displaymath}
\,\,\,\,\,\,\,\,\
C_{abcd}
= R_{abcd} +\frac{1}{2}(g_{ad}R_{cb}+g_{bc}R_{da}-g_{ac}R_{db}-g_{bd}R_{ca}) + 
  \frac{1}{6}(g_{ac}g_{db}-g_{ad}g_{cb})R .
\end{displaymath}
Both tensors $E_{ab}$ and $H_{ab}$ are spatial symmetric
tracefree rank-two tensors (irreducible tensors) \cite{Maar,Maar2,Maar3}. Each of them has 5
independent components: $E_{i}$ and $H_{i}$, $i=1,2,3,4,5$ respectively, which satisfy the
transformation properties of an irreducible rank-two tensor under the
discrete group operations $P$ and $T$. By using the transformation
properties table of section 5.1 for $j=2$, we have
\begin{displaymath}
\begin{array}{c|cc}
{\rm Discrete} & E_{i} & H_{i}  \\
{\rm Operation} & & \\   \hline
P & + & - \\
T & - & +
\end{array}
\end{displaymath}
Thus, the Weyl tensor splits irreducibly and covariantly into these
two symmetric tracefree tensors.

By using the matrix form for the irreducible tensors $E_{ab}$ and
$H_{ab}$ and the definition \cite{Maar} of the curl operator ($(\nabla \times T)_{ab}$)
in (\ref{b1}) we obtain two matrix equations. Each of them gives rise
to 5 independent equations. The 5 equations from the first one are
\begin{eqnarray}
-\frac{1}{2}\partial_{x}E_{3}+\frac{1}{2}\partial_{y}E_{2}+\partial_{z}E_{4}+\frac{1}{c}\frac{\partial H_{1}}{\partial t}=0 
 ,\nonumber \\
-\frac{1}{2}\partial_{x}(E_{4}+\sqrt{3}E_{5})-\frac{1}{2}\partial_{y}E_{1}+\frac{1}{2}\partial_{z}E_{3}+\frac{1}{c}\frac{\partial H_{2}}{\partial t}=0 
 ,\nonumber \\
\frac{1}{2}\partial_{x}E_{1}-\frac{1}{2}\partial_{y}(E_{4}-\sqrt{3}E_{5})-\frac{1}{2}\partial_{z}E_{2}+\frac{1}{c}\frac{\partial H_{3}}{\partial t}=0 
 ,\nonumber \\
\frac{1}{2}\partial_{x}E_{2}+\frac{1}{2}\partial_{y}E_{3}-\partial_{z}E_{1}+\frac{1}{c}\frac{\partial H_{4}}{\partial t}=0 
 ,\nonumber \\
\frac{\sqrt{3}}{2}\partial_{x}E_{2}-\frac{\sqrt{3}}{2}\partial_{y}E_{3}+\frac{1}{c}\frac{\partial H_{5}}{\partial t}=0
 ,
\label{b2}
\end{eqnarray}
and the 5 equations from the second one are
\begin{eqnarray}
-\frac{1}{2}\partial_{x}H_{3}+\frac{1}{2}\partial_{y}H_{2}+\partial_{z}H_{4}-\frac{1}{c}\frac{\partial E_{1}}{\partial t}=0 
 ,\nonumber \\
-\frac{1}{2}\partial_{x}(H_{4}+\sqrt{3}H_{5})-\frac{1}{2}\partial_{y}H_{1}+\frac{1}{2}\partial_{z}H_{3}-\frac{1}{c}\frac{\partial E_{2}}{\partial t}=0 
 ,\nonumber \\
\frac{1}{2}\partial_{x}H_{1}-\frac{1}{2}\partial_{y}(H_{4}-\sqrt{3}H_{5})-\frac{1}{2}\partial_{z}H_{2}-\frac{1}{c}\frac{\partial E_{3}}{\partial t}=0 
 ,\nonumber \\
\frac{1}{2}\partial_{x}H_{2}+\frac{1}{2}\partial_{y}H_{3}-\partial_{z}H_{1}-\frac{1}{c}\frac{\partial E_{4}}{\partial t}=0 
 ,\nonumber \\
\frac{\sqrt{3}}{2}\partial_{x}H_{2}-\frac{\sqrt{3}}{2}\partial_{y}H_{3}-\frac{1}{c}\frac{\partial E_{5}}{\partial t}=0
 .
\label{b3}
\end{eqnarray}
We have 10 equations in total. By comparing this set of equations
with the 10 equations of (\ref{eq:39})
that come from the constraint equation
for $j=2$, we find the components of the two oscillating interacting
fields $F$ and $G$ in terms of the components of the electric and magnetic Weyl
tensor $E$ and $H$. The relations are 
\begin{equation}
\begin{array}{cccccccc}
F_{i}=E_{i} ,& & & 
G_{i}=H_{i} , 
\end{array}
\label{b4}
\end{equation}
and they satisfy the transformation properties rules under $P$
and $T$ for $F$ and $G$ (table of section 5.1).

\section{Conclusion}

In the quantum description of the electromagnetic field, photons are
the fundamental building blocks.  Photons are described by a 4-vector
$k$ that obeys $k \cdot k=0$ in free space, and a helicity index
indicating a projection of an angular momentum $\pm 1$ (in units of $\hbar$) along the
direction of propagation of the photon.  Every physical state is
described by a superposition of the photon basis states, and every
superposition describes a possible physical state.  In this description
of the electromagnetic field no equations are necessary.

The classical description of the electromagnetic field proceeds
along somewhat different lines.  A multicomponent field $(\vec{E}, \vec{B})$
is introduced at each point in space-time.  The components of the
field (tensor) transform in a very elegant way under homogeneous
Lorentz transformations.  If the field is Fourier transformed from the
coordinate to the momentum representation, then each 4-momentum has
six components associated with it.  Since the quantum description has
only two independent components associated with each 4-momentum, there
are four dimensions worth of linear combinations of the classical
field components that do not describe physically allowed states, for
each 4-momentum.  Some mechanism must be derived for annihilating
these non-physical superpositions.  This mechanism is the set of equations
discovered by Maxwell for the electromagnetic field in the absence of sources. 
Similar equations hold for gravitational radiation in Minkowski spacetime.
In this sense, these equations are an
expression of our ignorance.

Group theory, by pointing to the appropriate
Hilbert space for the electromagnetic field, allows us to relate physical states to
arbitrary superpositions of basis states.  Since no superpositions are
forbidden, no equations are necessary.

We have derived constraint equations on the $2(2j+1)$ components
of a manifestly covariant field transforming under the representation
$D^{j0+0j}$ of the ILG.  There are two constraint equations:

\begin{equation}
\begin{array}{ccccc}
{\rm curl}(T_{E}^{(j)} + iT_{B}^{(j)} ) & - &  \frac{i}{c} \frac{\partial}{\partial t}(T_{E}^{(j)} + iT_{B}^{(j)}) &=&0 , \\
 & & & & \\
 & & {\rm div}(T_{E}^{(j)} + iT_{B}^{(j)}) & = &0 . \end{array}
\end{equation}
For $j=1$ these are the electric and magnetic fields.
For $j=2$ these are the second order tracefree symmetric 
gravito-electric and gravito-magnetic field tensors that are projected 
from  the Weyl tensor.

It is remarkable that gravitational radiation in flat space-time can be 
described in essentially the same way as Maxwell's classical equations.

\section{Acknowledgment}

We would like to thank Dr. R. Jantzen and Dr. B. Mashhoon for helpful comments on this work.

\end{document}